\documentclass[journal,final,draftcls,letterpaper,oneside,twocolumn,romanappendices]{IEEEtran}
\IEEEoverridecommandlockouts
\usepackage{silence} 
\usepackage[T1]{fontenc}
\usepackage[utf8]{inputenc}
\usepackage{amsmath,amsfonts,amsthm,amssymb}
\usepackage{algorithmic,algorithm}
\usepackage{array,stackrel}
\usepackage[caption=false,font=normalsize,labelfont=sf,textfont=sf]{subfig}
\usepackage{textcomp}
\usepackage{stfloats}
\usepackage{url}
\usepackage{verbatim}
\usepackage{graphicx}
\usepackage{cite}
\usepackage{xcolor}
\usepackage{pdfcolmk}
\usepackage{refcount}
\usepackage{esint}
\usepackage{hhline}
\usepackage{bm}
\usepackage{xfrac}
\usepackage{makecell}
\usepackage{float}
\usepackage{soul}
\usepackage{lipsum}  
\usepackage[switch]{lineno}

\usepackage{cases}
\usepackage[]{footmisc}
\def\BibTeX{{\rm B\kern-.05em{\sc i\kern-.025em b}\kern-.08em
    T\kern-.1667em\lower.7ex\hbox{E}\kern-.125emX}}

\theoremstyle{plain}
\newtheorem{thm}{\protect\theoremname}
\theoremstyle{plain}

\usepackage{stackengine}

\usepackage[caption=false,font=normalsize,labelfont=sf,textfont=sf]{subfig}

\makeatother

\providecommand{\lemmaname}{Lemma}
\providecommand{\theoremname}{Theorem}

\theoremstyle{definition}

\usepackage{tabularx,booktabs}
\newcolumntype{C}{>{\centering\arraybackslash}X} 
\usepackage{booktabs}
\PassOptionsToPackage{normalem}{ulem}
\usepackage{ulem}
\usepackage[unicode=true,bookmarks=true,bookmarksnumbered=true,bookmarksopen=true,bookmarksopenlevel=1,
 breaklinks=false,pdfborder={0 0 0},pdfborderstyle={},backref=false,colorlinks=true, urlcolor=blue, linkcolor=blue, citecolor=blue]{hyperref}
 \hypersetup{pdftitle={Your Title},
 pdfauthor={Your Name},
 pdfpagelayout=OneColumn, pdfnewwindow=true, pdfstartview=XYZ, plainpages=false}
\PassOptionsToPackage{unicode}{hyperref}
\PassOptionsToPackage{naturalnames}{hyperref}
\usepackage[cmintegrals]{newtxmath}
\usepackage[caption=false,font=normalsize,labelfont=sf,textfont=sf]{subfig}
\newcounter{MYtempeqncnt}

\makeatletter 
\usepackage{orcidlink}
\begin{document}

\title{IRS Compensation of Hyper-Rayleigh Fading: \\How Many Elements Are Needed?
}

\author{Aleksey S.~Gvozdarev\,\orcidlink{0000-0001-9308-4386},~\IEEEmembership{Member,~IEEE}
\thanks{The author is with the Department of Intelligent Radiophysical Information Systems (IRIS), P. G. Demidov Yaroslavl State University, Yaroslavl, 150003, Russia (e-mail: \href{mailto:asg.rus@gmail.com}{asg.rus@gmail.com})}
\thanks{This work was supported by Russian Science Foundation under Grant 24-29-00516 (\href{https://rscf.ru/en/project/24-29-00516/}{https://rscf.ru/en/project/24-29-00516/}).}%
\thanks{\copyright 2026 IEEE.  Personal use of this material is permitted.  Permission from IEEE must be obtained for all other uses, in any current or future media, including reprinting/republishing this material for advertising or promotional purposes, creating new collective works, for resale or redistribution to servers or lists, or reuse of any copyrighted component of this work in other works.}
\thanks{Digital Object Identifier: 10.1109/LWC.2026.3656740}
}

\maketitle

\begin{abstract}
The letter introduces and studies the problem of defining the minimum number of Intelligent Reflecting Surface (IRS) elements needed to compensate for heavy fading conditions in multipath fading channels. The fading severity is quantified in terms of Hyper-Rayleigh Regimes (HRRs) (i.e., full-HRR (worst-case conditions), strong-, weak-, and no-HRR), and the channel model used (Inverse Power Lomax (IPL)) was chosen since it can account for all HRRs. The research presents the derived closed-form channel coefficient envelope statistics for the single IRS-element channel with IPL statistics in both subchannels and total IRS-assisted channel, as well as tight approximations for the channel coefficient and instantaneous signal-to-noise ratio (SNR) statistics for the latter. The derived expressions helped estimate channel parameters corresponding to the specific HRRs of the total channel and demonstrate that while both single links (i.e., ``source-IRS'' and ``IRS-destination'') are in full-HRR, the minimum number of IRS elements needed to bring the total IRS-assisted link (``source-IRS-destination'') out of full-HRR is no less than $6$ (for the whole range on the IPL scale parameter corresponding  full-HRR). Furthermore, the minimum number of IRS elements required to bring the total IRS-assisted link into no-HRR is $14$ (under the same conditions).

\end{abstract}


\section{Introduction}\label{S-1}

Intelligent Reflecting Surfaces (IRS) have emerged as a technology of paramount importance for future wireless communications, with strong potential for integration into 6G standards \cite{ITU22}. Their ability to dynamically manipulate electromagnetic wave propagation makes them a powerful tool for enhancing signal coverage, spectral efficiency, and energy efficiency in next-generation networks \cite{Dir19}. IRS systems typically consist of numerous passive reflecting elements, but face key challenges in supporting multiple users and dynamically adjusting reflective properties in real time \cite{Dir19}. This is often addressed by subdividing the IRS panel into sub-arrays for each user group \cite{Zha25}, raising a critical research question: What is the minimum viable sub-array size for effective IRS operation? The answer depends on conflicting trade-offs \cite{Zap21, Li22, Cha25}: fewer elements reduce reflection efficiency but enable faster adaptation and support for more users.

While classical approaches model IRS sub-blocks as antenna arrays deriving the minimum number of elements based on  beamforming parameters \cite{Han21} (e.g., directivity, beamwidth, and spatial resolution, etc.), or focuses on coverage/rate/energy efficiency adaptaion, this work adopts a fundamentally different methodology focusing on fading mitigation (i.e., fading-centric). We specifically address severe fading conditions through the framework of hyper-Rayleigh analysis \cite{Gar19}, which quantifies fading severity via Hyper-Rayleigh Regimes (HRRs). This leads to our fundamental research questions: When each subchannel operates in worst-case (full-HRR) conditions, can the total IRS-assisted link be brought out of full-HRR, and if so, what is the minimum number of IRS elements required? Furthermore, what is required to achieve no-HRR performance under these challenging conditions?

The presented research addresses the questions raised. To derive quantitative results, the research adopts a specific fading channel model, i.e., the inverse power Lomax model \cite{Gvo24}, suitable for wireless communications (demonstrated for various communication systems), which is capable of handling heavy fading conditions. The major contributions of this work can be summarized as follows: \textit{a)} for the single IRS-element channel with IPL statistics in both subchannels, the probability density function (PDF) and moment-generating function (MGF) of the channel coefficient envelope are derived in closed form;
\textit{b)} for the total IRS-assisted channel, exact expression for the PDF of the channel coefficient and its tight approximations are derived;
\textit{c)} for the total IRS-assisted channel, the conditions for the specific hyper-Rayleigh regimes (i.e., no-HRR, weak-HRR, strong-HRR, and full-HRR) are derived in closed form;
\textit{d)} the minimum number of IRS elements needed either to bring the total IRS-assisted link out of the full-HRR or drive it into the no-HRR.

\section{Preliminaries}\label{S-2}
\subsection{IRS-assisted communication model}\label{Ss-2-1}
Let us assume the classical communication model between the source (S) and destination (D) (with no direct Line-of-Sight) through the IRS panel of $N_{IRS}$ elements. Since the effects of adaptation are of no interest for the current research, hereafter one assumes a perfectly tuned IRS, which completely compensates phase shifts of the waves impinging on the panel and reflected waves to optimize some chosen performance metrics at the destination point \cite{Dir19}. Denoting the channel complex coefficients in the link between the source and \textit{j}-th IRS element ($(\text{S})\to(\text{IRS}_j)$) as $\dot h_s$ and in the remaining link (i.e., $(\text{IRS}_j)\to(\text{D})$) as $\dot h_d$, the total coefficient of the effective channel is given by $ h_\Sigma=\sum_{j=1}^{N_{IRS}}|\dot h_{s_j}||\dot h_{d_j}|$. In the presence of fading, $h_\Sigma$ is a random variable completely defined by the probability density distributions of $|\dot h_{s_j}|$ and $|\dot h_{d_j}|$. To quantify the fading severity, as was mentioned earlier, one can resort to the so-called hyper-Rayleigh regimes~\cite{Gar19,Gvo23}.

\subsection{Hyper-Rayleigh fading classification}\label{Ss-2-2}

The classical definition of HRRs (introduced in \cite{Gar19}) is as follows: the channel exhibits hyper-Rayleigh behavior if either the amount of fading (AoF), outage probability (OP), or ergodic capacity (EC) is worse than that of a Rayleigh channel. If all of these conditions are met, the regime is called full-HRR; if any two are met -- strong-HRR; if one -- weak-HRR; otherwise, no-HRR is declared. The first and the last are of specific interest for the current research, since they describe the worst-case and best-case propagation conditions. Recalling the definition of AoF  (i.e., $\text{AoF}=\frac{\mathbb{E}\{\gamma^2\}-\left(\mathbb{E}\{\gamma\}\right)^2}{\left(\mathbb{E}\{\gamma\}\right)^2}$), the representation of OP in terms of the power offset (PO) $\Delta_{PO}$ and diversity gain (DG) $\mathrm{G}_D$ \cite{Gar19}, capacity loss metric $\Delta\overline{\mathrm{C}}=-G_e+\mathbb{M}'(0)$ (here $G_e$ is the Euler-Mascheroni constant and $\mathbb{M}(r)=\frac{\mathbb{E}\{\gamma^r\}}{\left(\mathbb{E}\{\gamma\}\right)^r}$ is the normalized \textit{r}-th moment), and corresponding values for the Rayleigh channel (serving as a benchmark), the HRR conditions can be represented as
\begin{equation}\label{eq-2.B-1}
\text{AoF}>1, \quad \mathrm{G}_D>1\cap \left(
\displaystyle \mathrm{G}_D=1 \cup\Delta_{PO}>1\\
\right),\quad
\displaystyle \Delta\overline{\mathrm{C}}>0.
\end{equation}

It must be specifically pointed out that the equality sign in \eqref{eq-2.B-1} (i.e., $\text{AoF}=1$, $\mathrm{G}_D=1$, $\Delta_{PO}=1$, $\Delta\overline{\mathrm{C}}=0$) directly corresponds to the Rayleigh fading. Since the evaluation of HRRs requires channel statistics (instantaneous SNR distribution), one has to define the corresponding channel model.

\subsection{Fading channel model}\label{Ss-2-3}

The research adopts Inverse Power Lomax distribution for both segments (i.e., $(\text{S})\to(\text{IRS}_j)$ and $(\text{IRS}_j)\to(\text{D})$) of the total channel. 
IPL model \cite{Gvo24} is a two-parametric one, described by scale and shape parameters $\alpha, \beta$ with the probability density function (PDF) of the channel coefficient envelope $|\dot h_i|$ (with $i=\{s;d\}$)  given by:
\begin{IEEEeqnarray}{rCl}
&&f_{|\dot h_i|}(|\dot h_i|)=2\frac{\alpha_i\beta_i}{\mho_i}\frac{\Omega_i^{\beta_i}}{|\dot h_i|^{2\beta_i+1}}\left(1+\frac{1}{\mho_i}
\left(\frac{\Omega_i}{|\dot h_i|^2}\right)^{\beta_i}\right)^{-(\alpha_i+1)},\label{eq-2.C-2}
\end{IEEEeqnarray}
where $\Omega_i=\mathbb{E}\{|\dot h_i|^2\}$, $\bar{\gamma}_i$ is the average SNR, $\mho_i=\left(\frac{\Gamma\left(1-\frac{1}{\beta_i}\right)\Gamma\left(\alpha_i+\frac{1}{\beta_i}\right)}{\Gamma(\alpha_i)}\right)^{\beta_i}$, and $\Gamma(\cdot)$ is the gamma-function \cite{DLMF}.
The choice of the model was motivated not only because it can hold all the HRRs, but also by the fact that it was verified on various experimental data \cite{Gvo24} (including, near-field Device-to-Device (D2D) and long-range cellular communications, indoor and outdoor, with line-of-sight and without, etc.) and demonstrated good agreement.  Such severe fading is typical in environments like urban canyons, industrial indoor settings, or near-field D2D communications, where the signal experiences deeper and more frequent fades than predicted by the Rayleigh model. Referring to \cite{Gvo24}, one can see that each of the channels exhibits full hyper-Rayleigh fading if $\alpha_i\beta_i=1$ and $\alpha_i<0.316$.  Thus, these conditions will be used later on for defining the required number of IRS elements.

\subsection{Problem statement}\label{Ss-2-4}
Summarizing all of the above, assuming that each channel (i.e., $(\text{S})\to(\text{IRS}_j)$ and $(\text{IRS}_j)\to(\text{D})$) exhibits full hyper-Rayleigh fading, this letter aims to answer two main questions:
\begin{itemize}
    \item What is the minimum number of IRS elements ($N^{*}_f$) needed to take the IRS-assisted system out of the full hyper-Rayleigh regime, i.e., $N^*_f~=~\underset{N_{IRS}}{\arg\min}\left\{\left(\mathrm{G}_D>1 \cap [\mathrm{G}_D=1\cup\Delta_{PO}<1]\right)
       \cap \right.$ \\ \mbox{\hspace{3.7cm}}$\left.\cap\mathrm{AoF}<1\cap\Delta\bar{\mathrm{C}}<0\right\}.$
    \item  What is the minimum number of IRS elements ($N^{*}_n$) required to elevate the entire link to no-HRR despite full-HRR conditions in the subchannels, i.e., \\$N^*_n=\underset{N_{IRS}}{\arg\min}\left\{\left(\mathrm{G}_D>1 \cap [\mathrm{G}_D=1\cup\Delta_{PO}<1]\right)\cup\right.$\mbox{\hspace{0pt}} \\  
    \mbox{\hspace{3.7cm}} $\cup  \left.\mathrm{AoF}<1\cup\Delta\bar{\mathrm{C}}<0\right\}.$
\end{itemize}

Formalizing the problem, with the system and channel models at hand, one can evaluate $N^*_f$ and $N^*_n$.

\section{Derived solution}\label{S-3}

To solve the stated above problem, one assumes a three-stage procedure: \textit{1)} derive the statistics of a single-IRS channel (i.e., find the distribution of $|\dot h_{s_j}||\dot h_{d_j}|$), \textit{2)} derive the statistics of a total-IRS channel (i.e., the distribution of $h_\Sigma$), \textit{3)} evaluate the conditions for specific HRRs of the  total-IRS channel and find (analytically of numerically) $N^*_f$ and $N^*_n$. 

\subsection{Closed-form IRS output envelope statistics}\label{Ss-3-1}

On the first step, let us denote $Z_j=|\dot h_{s_j}||\dot h_{d_j}|$, then the following result for the IRS link with a single IRS element holds.  

\begin{thm} The probability density function of the single IRS-element channel coefficient $Z_j$ is given by:
\begin{IEEEeqnarray}{rCl}\label{eq-3.A-1}
&&f_{Z_j}(z_j)=\frac{\mathrm{H}_{2,2}^{\,2,2}\!\left(\left.
\sqrt{\frac{\mho_s^{\frac{1}{\beta_s}}\mho_d^{\frac{1}{\beta_d}}}{\Omega_s\Omega_d}}z_j\right|\,{\begin{matrix}\left(0,\frac{1}{2\beta_s}\right),\left(0,\frac{1}{2\beta_d}\right)\\ \left(\alpha_s,\frac{1}{2\beta_s}\right),\left(\alpha_d,\frac{1}{2\beta_d}\right)\end{matrix}}\; 
\right)}{z_j\Gamma(\alpha_s)\Gamma(\alpha_d)},
\end{IEEEeqnarray}
where $\mathrm{H(\cdot)}$ is the Fox H-function (see \cite[Chapter 1.1, eq. (1.1.1)-(1.1.6)]{Kil04}.
\end{thm}
\begin{IEEEproof}
To prove Theorem~1, one can rewrite \eqref{eq-2.C-2} in hypergeometric form recalling that 
$\frac{\left(1+z\right)^{-(\alpha_i+1)}}{\Gamma(\alpha_i+1)}=\mathrm{H}_{1,1}^{\,1,1}\!\left(\left.
z\right|\,{\begin{matrix}(-\alpha_i,1)\\ (0,1)\end{matrix}}\; 
\right)$, where $\dot h_{i_j}$, where $\dot h_{i_j}$ is the transmission coefficient for the \textit{i}-th channel ($i=\{s;d\}$), communicating through the \textit{j}-th IRS element. Thus, combining this representation with the product $Z_j$ distribution $f_{Z_j}(z_j)=\int_0^{\infty} f_{|\dot h_{s_j}|}(x)f_{|\dot h_{d_j}|}\left(\frac{z_j}{x}\right)\frac{{\rm d}x}{x}$ having applied \cite[Theorem 2.9, eq. (2.8.4)]{Kil04} and  \cite[Property 2.5, eq. (2.1.5)]{Kil04}, \eqref{eq-3.A-1} follows.
\end{IEEEproof}

On the second step, let us assume that the angular size of the $N_{IRS}$-element sub-block is small compared to source and destination distances, which means $\forall j=1\ldots N_{IRS}: \alpha_{s_j}=\alpha_{s}, \alpha_{d_j}=\alpha_{d}, \beta_{s_j}=\beta_{s}, \beta_{d_j}=\beta_{d}$.  Then, with the help of Theorem~1, one can derive the statistics of $h_\Sigma$ in closed form as follows.
\begin{thm} The probability density function of the output envelope $f_{h_\Sigma}(h_\Sigma)$ of the IRS-assisted  model with $N_{IRS}$ elements can be implicitly represented as:
\begin{IEEEeqnarray}{rCl}\label{eq-3.A-2}
f_{h_\Sigma}(h_\Sigma)&=&\mathcal{L}^{-1}\left\{\left(\mathcal{M}_Z(s)\right)^{N_{IRS}}\right\},
\end{IEEEeqnarray}
where $\mathcal{L}^{-1}\left\{\cdot\right\}$ is the inverse Laplace transform of the moment generating function $\mathcal{M}_Z(s)$ of the single IRS-element channel coefficient evaluated in closed form as follows:
\begin{IEEEeqnarray}{rCl}\label{eq-3.A-3}
&&\hspace{-10pt}\mathcal{M}_Z(s)=\frac{\mathrm{H}_{3,2}^{\,2,3}\!\left(\left.
\sqrt{\frac{\mho_s^{\frac{1}{\beta_s}}\mho_d^{\frac{1}{\beta_d}}}{\Omega_s\Omega_d}}s^{-1}\right|\,{\begin{matrix}\left(1,1\right),\left(0,\frac{1}{2\beta_s}\right),\left(0,\frac{1}{2\beta_d}\right)\\ \left(\alpha_s,\frac{1}{2\beta_s}\right),\left(\alpha_d,\frac{1}{2\beta_d}\right)\end{matrix}}\; 
\right)}{\Gamma(\alpha_s)\Gamma(\alpha_d)}.
\end{IEEEeqnarray}

The explicit form of $f_{h_\Sigma}(h_\Sigma)$ is given by \eqref{eq-3.A-2-1} (at the bottom of the page).

\begin{figure*}[!b]
\hrulefill
\normalsize
\setcounter{MYtempeqncnt}{0}
\setcounter{equation}{7}
\vspace*{-4pt}
\begin{IEEEeqnarray}{rCl}\label{eq-3.A-2-1}
&&f_{h_\Sigma}(h_\Sigma)=\frac{h_\Sigma^{-1}}{\left(\Gamma(\alpha_s)\Gamma(\alpha_d)\right)^{N_{IRS}}}{\rm H}_{\scriptsize{0,1}:{ \underbrace{3,2;\ldots;3,2}_{\scriptsize {N_{IRS}}-times}}}^{\scriptsize{0,0}:{\overbrace{ 2,3;\ldots;2,3}^{\scriptsize {N_{IRS}}-times}}} 
\left[ \begin{matrix}{\sqrt{\frac{\mho_s^{\frac{1}{\beta_s}}\mho_d^{\frac{1}{\beta_d}}}{\Omega_s\Omega_d}}h_\Sigma} \\ \vdots \\ {\sqrt{\frac{\mho_s^{\frac{1}{\beta_s}}\mho_d^{\frac{1}{\beta_d}}}{\Omega_s\Omega_d}}h_\Sigma} 
\end{matrix} 
\; \vrule\; \begin{matrix} {}\\ {} \\ {} \end{matrix}
\begin{matrix} \mbox{---} \\ {\vdots} \\ {\left( 1; 1\ldots 1\right)} \end{matrix} 
\begin{matrix} {:} \\ { } \\ {:} \end{matrix} 
\begin{matrix} { \left(0,\frac{1}{2\beta_s}\right); \ldots; \left(0,\frac{1}{2\beta_s}\right); \left(0,\frac{1}{2\beta_d}\right); \ldots; \left(0,\frac{1}{2\beta_d}\right);(1,1)} \\ {\vdots} \\ {\left(\alpha_s,\frac{1}{2\beta_s}\right); \ldots; \left(\alpha_s,\frac{1}{2\beta_s}\right);\left(\alpha_d,\frac{1}{2\beta_d}\right); \ldots; \left(\alpha_d,\frac{1}{2\beta_d}\right)} \end{matrix}
\right].\IEEEnonumber\\
\end{IEEEeqnarray}
\end{figure*}

\end{thm}
\begin{IEEEproof}
Assuming the independence of the $Z_j$ it is convenient to describe the statistics of $h_\Sigma$ in terms of their MGFs since $\mathcal{M}_{h_\Sigma}(s)=\left(\mathcal{M}_{Z_j}(s)\right)^{N_{IRS}}$. Furthermore, the PDF $f_{H_\Sigma}(h_\Sigma)$ can be recovered as the inverse Laplace transform of $\mathcal{M}_{h_\Sigma}(s)$. Thus, the last missing piece  required to prove Theorem~2 is the MGF $\mathcal{M}_{Z_j}(s)$. It can be evaluated by relating the definition of the MGF with Laplace transform $\mathcal{M}_{Z_j}(-s)=\mathbb{E}\{e^{-s Z_j}\}=\mathcal{L}\{f_{Z_j}(z_j);s\}$ and applying \cite[Corollary 2.3.1, eq. (2.5.25)]{Kil04}. After some mathematical simplifications, \eqref{eq-3.A-3} follows. To complete the proof of Theorem~2 one can expand the MGF of total channel in terms of $N_{IRS}$-fold contour integrals:
\begin{IEEEeqnarray}{rCl}\label{eq-3.A-4}
&&\hspace{-5pt}\mathcal{M}_{h_\Sigma}(s)=\frac{\left(\Gamma(\alpha_s)\Gamma(\alpha_d)\right)^{-N_{IRS}}}{(2\pi i)^{N_{IRS}}} \int_{\mathcal{H}_1} \ldots \int_{\mathcal{H}_{N_{IRS}}}s^{\sum_{l=1}^{N_{IRS}} \sigma_l} \times \IEEEnonumber\\
&&\hspace{10pt}\times\prod_{l=1}^{N_{IRS}}\Gamma(\sigma_l)\Gamma\left(1+\frac{\sigma_l}{\beta_{s_l}}\right)\Gamma\left(\alpha_{s_l}-\frac{\sigma_l}{\beta_{s_l}}\right)\Gamma\left(1+\frac{\sigma_l}{\beta_{d_l}}\right)\times \IEEEnonumber\\
&& \hspace{10pt}\times \Gamma\left(\alpha_{d_l}-\frac{\sigma_l}{\beta_{d_l}}\right)\left(\frac{\mho_s^{\frac{1}{\beta_s}}\mho_d^{\frac{1}{\beta_d}}}{\Omega_s\Omega_d}\right)^{-\sum_{l=1}^{N_{IRS}} \frac{\sigma_l}{2}}{\rm d}\sigma_1\ldots{\rm d}\sigma_l.
\end{IEEEeqnarray}

Recall that $f_{h_\Sigma}(h_\Sigma)=\mathcal{L}^{-1} \{ \mathcal{M}_{h_\Sigma} (-s); h_\Sigma \}$.
Taking the inverse Laplace transform  of \eqref{eq-3.A-4} and applying the definition of multivariate Fox H-function (see \cite[Appendix A]{Mat09}) finalizes the proof of Theorem~2.
\end{IEEEproof}

It should be pointed out that in the current research (for convenience only), the author confines oneself with analytically tractable results (i.e., intensionally avoiding multivariate hypergeometric-type functions). Thus, even though \eqref{eq-3.A-3} presents the exact solution, its computational complexity is too high for large $N_{IRS}$. At the same time, although \eqref{eq-3.A-2} cannot be assumed as a closed-form representation, if needed, it can be efficiently computed numerically via the procedures, exhaustively discussed in \cite{Coh07}.

\subsection{Approximate IRS output SNR statistics}\label{Ss-3-2}

Nevertheless, since further practical implementation of the derived results (e.g., error rate, outage, or capacity analysis) requires some closed-form expression for the PDF \eqref{eq-3.A-2}, its approximation is a clear and straightforward solution. Thus, the following statements present the approximate closed-form expression for the PDF of the output envelope $f_{h_\Sigma}(h_\Sigma)$ and (what is more importantly from the practical perspectives) output SNR $f_{\gamma_\Sigma}(\gamma_\Sigma)$. 
\begin{thm} The probability density function of the output envelope $f_{h_\Sigma}(h_\Sigma)$ and instantaneous SNR $f_{\gamma_\Sigma}(\gamma_\Sigma)$ of the IRS-assisted  model with $N_{IRS}$ elements can be tightly approximated as:
\begin{IEEEeqnarray}{rCl}
f_{\gamma_\Sigma}(\gamma_\Sigma)&=&\frac{\left(\hat\alpha _g \left(\hat\alpha _g+1\right)\frac{\gamma_\Sigma}{\bar{\gamma }_{\Sigma }}\right){}^{\frac{\hat\alpha _g}{2}}}{2 \gamma _{\Sigma } \Gamma \left(\hat\alpha _g\right)}e^{-\sqrt{\hat\alpha _g \left(\hat\alpha_g+1\right)\frac{\gamma _{\Sigma }}{\bar{\gamma }_{\Sigma }}}},\label{eq-3.B-2}
\end{IEEEeqnarray}
where $\bar\gamma_\Sigma$ is the average SNR and $\hat\alpha_g$ is explicitly given by
\begin{IEEEeqnarray}{rCl}\label{eq-3.B-4}
&&\hat\alpha_g= \frac{N_{IRS}}{\frac{\Gamma\left(\alpha_s+\frac{2}{\beta_s}\right)\Gamma\left(1-\frac{2}{\beta_s}\right)\Gamma\left(\alpha_d+\frac{2}{\beta_d}\right)\Gamma\left(1-\frac{2}{\beta_d}\right)\Gamma(\alpha_s)\Gamma(\alpha_d)}
{\left(\Gamma\left(\alpha_s+\frac{1}{\beta_s}\right)\Gamma\left(1-\frac{1}{\beta_s}\right)\Gamma\left(\alpha_d+\frac{1}{\beta_d}\right)\Gamma\left(1-\frac{1}{\beta_d}\right)\right)^2}-1}.
\end{IEEEeqnarray}
\end{thm}

\begin{IEEEproof}
To prove \eqref{eq-3.B-2}, first one can note that $Z_j>0$, so Gaussian-type approximations \cite{Dir19} will tend to overestimate the left tail of the PDF, which describes heavy fading (i.e., full- and strong-HRRs) and thus is of primary importance herein. To take this issue into account, one resorts to the gamma-type approximation (see \cite[Section 2.2.2]{Pri04}): $f_{h_\Sigma}(h_\Sigma)=\frac{h_{\Sigma}^{\alpha_G-1}}{\Gamma(\alpha_G)\beta_G^{\alpha_G}}e^{-\frac{h_{\Sigma}}{\beta_G}}$, with $\alpha_G=\frac{\left(\mathbb{E}\{h_\Sigma\}\right)^2}{\mathbb{E}\{h_\Sigma^2\}-\left(\mathbb{E}\{h_\Sigma\}\right)^2}$, and $\beta_G=\frac{\mathbb{E}\{h_\Sigma^2\}-\left(\mathbb{E}\{h_\Sigma\}\right)^2}{\mathbb{E}\{h_\Sigma\}}$, which is much simpler than the squared $K_G$ approximation, presented in \cite{Yan20}. So, first two moments provide the explicit expression for $f_{h_\Sigma}(h_\Sigma)$ with $\alpha_G=\frac{N_{IRS}\left(\mathbb{E}\{Z_j\}\right)^2}{\mathbb{E}\{Z_j^2\}-\left(\mathbb{E}\{Z_j\}\right)^2}$, $ \beta_G=\frac{\mathbb{E}\{Z_j^2\}-\left(\mathbb{E}\{Z_j\}\right)^2}{\mathbb{E}\{Z_j\}}$, which can be evaluated by applying \cite[Theorem 2.2, eq. (2.5.7)]{Kil04} to the derived PDF \eqref{eq-3.A-1}, yielding: $\mathbb{E}\{Z_j^k\}=\frac{\Gamma\left(\alpha_s+\frac{k}{2\beta_s}\right)\Gamma\left(\alpha_d+\frac{k}{2\beta_d}\right)}{\left(\Gamma\left(1-\frac{k}{2\beta_s}\right)\Gamma\left(1-\frac{k}{2\beta_d}\right)\right)^{-1}}
\frac{\left(\Gamma\left(\alpha_s\right)\right)^{-1}}{\Gamma\left(\alpha_d\right)}\left(\frac{\mho_s^{\frac{1}{\beta_s}}\mho_d^{\frac{1}{\beta_d}}}{\Omega_s\Omega_d} \right)^{\frac{k}{2}}$,
with the limitations, implied by the performed transformations, i.e., it is valid for $-2\alpha_s\max(\beta_s,\beta_d)<k<2\alpha_s\min(\beta_s,\beta_d)$.

To finalize the proof, recall that for gamma distribution $\Omega_h=\mathbb{E}\{h_{\Sigma}^2\}=\alpha_G(\alpha_G+1)\beta_G^2$ and the envelope $h_{\Sigma}$ is related to the instantaneous SNR $\gamma_{\Sigma}$ as $\frac{h_{\Sigma}^2}{\Omega_h}=\frac{\gamma_{\Sigma}}{\bar\gamma_{\Sigma}}$, performing random variable transformation (i.e., $f_{\gamma_\Sigma}({\gamma_\Sigma})=\frac{1}{2}\sqrt{\frac{\Omega_h}{{\gamma_\Sigma}{\bar\gamma_\Sigma}}}f_{h_\Sigma}\left(\sqrt{\frac{\Omega_h}{{\bar\gamma_\Sigma}}{\gamma_\Sigma}}\right)$); after some simplifications, \eqref{eq-3.B-2} follows.
\end{IEEEproof}

To verify the approximation quality, numerical simulations were performed\footnote{For detailed results' description and executable simulation files, please, see https://doi.org/10.5281/zenodo.17487676} and the obtained results demonstrate that the  approximation is surprisingly tight even for a small number of IRS elements (e.g., $3$ or $4$), which makes it suitable for further HRR analysis.  %

\subsection{Hyper-Rayleigh regimes of the IRS-assisted channel}\label{Ss-3-3}

Capitalizing on the derived PDFs, the boundaries of each hyper-Rayleigh regime can be identified explicitly.

\begin{thm}\label{thm:4} The IRS-assisted communication system exhibits full-HRR if $\hat\alpha_g\leq 2$, strong-HRR if $2<\hat\alpha_g\lessapprox G_e^{-1}(1+\sqrt{1-\frac{G_e}{3}})$, weak-HRR if $G_e^{-1}(1+\sqrt{1-\frac{G_e}{3}})<\hat\alpha_g\leq \frac{1}{2}(3+\sqrt{33})$, and no-HRR otherwise. Moreover, the strongest requirements are implied by the amount of fading; the weakest -- by the outage probability.
\end{thm}
\begin{IEEEproof}
    To prove the statement, one has to evaluate AoF and asymptotic OP and EC (for $\bar\gamma_\Sigma\to\infty$). Noting that \eqref{eq-3.B-2} is a generalized gamma distribution (see \cite[Section 17.8.7]{Joh94}) with shape parameters $\frac{\hat\alpha_g}{2}, \frac{1}{2}$ and scale parameter $\frac{\bar{\gamma }_{\Sigma }}{\hat\alpha_g(\hat\alpha_g+1)}$, the moments (for AoF and capacity loss metric) can be evaluated via \cite[eq. 17.117]{Joh94}. Thus, $\text{AoF}=\frac{2}{\hat\alpha_g}\frac{(2\hat\alpha_g+3)}{(\hat\alpha_g+1)}$ and $\Delta\overline{\mathrm{C}}=-G_e+\ln(\hat\alpha_g(\hat\alpha_g+1))-2\psi^{(0)}(\hat\alpha_g)$, where $\psi^{(0)}(\cdot)$ is the digamma function \cite{DLMF}. The high-SNR asymptotics of the OP can be obtained in the following form: $\mathrm{P_{out}}\sim \frac{(\hat\alpha_g(\hat\alpha_g+1))^{\frac{\hat\alpha_g}{2}}}{\Gamma(\hat\alpha_g+1)}\left(\frac{\gamma_{th_\Sigma}}{\bar\gamma_\Sigma}\right)^{\frac{\hat\alpha_g}{2}}$, from where one can deduce that $\mathrm{G}_D=\frac{\hat\alpha_g}{2}$, and $\Delta_{PO}=\frac{(\hat\alpha_g(\hat\alpha_g+1))^{\frac{\hat\alpha_g}{2}}}{\Gamma(\hat\alpha_g+1)}$. Applying the HRR definitions \eqref{eq-2.B-1}, it should be noted that the restrictions on $\hat\alpha_g$ for AoF and OP can be evaluated explicitly (yielding the strongest and the weakest constraints respectively), but for EC only numerical solution can be sought. Although, a good approximation (with relative error less than $2\%$) of the latter can be obtained by using 2-order Taylor expansion of $\Delta\overline{\mathrm{C}}$, which yields $G_e^{-1}\left(1+\sqrt{1-\frac{G_e}{3}}\right)\approx 3.29$ (with the exact value around $3.33$), finalizing the proof.
\end{IEEEproof}

\subsection{Derived minimum number of required IRS elements}\label{Ss-3-4}

Since the HRRs' boundaries are completely defined by Theorem~4, their combination with \eqref{eq-3.B-4} gives the exact values for $N^*_f, N^*_n$, although, due to the intricate dependence of $\hat\alpha_g$ on the channel parameters ($\alpha_s, \alpha_d, \beta_s, \beta_d$) no exact solutions can be derived.
Hence, further (solemnly for illustrative purposes) one assumes the case of equal channels, (i.e., $\alpha_s=\alpha_d=\alpha_0, \beta_s=\beta_d=\beta_0$). Since the we are interested in full-HRR, which is attained for $\alpha_0\beta_0<1$ and $0<\alpha_0<0.316$, for later on one assumes that $\alpha_0=\frac{1}{2\beta_0}$ and $0<\alpha_0<0.316$.

\begin{thm} For the IRS-assisted communication system with identical IPL  fading channels in both links (i.e., $(\text{S})\to(\text{IRS}_j)$ and $(\text{IRS}_j)\to(\text{D})$) experiencing full-HRR each (i.e.,  $\alpha_0=\frac{1}{2\beta_0}$ and $0<\alpha_0<0.316$), the number of IRS elements required to elevate the total channel out of full-HRR or bring it to the no-HRR is $N^*_f=2$, $N^*_n=4$ for $\alpha_0\to 0$ and $N^*_f=6$, $N^*_n=14$ for $\alpha_0\to 0.316$.
\end{thm}

\begin{IEEEproof}
    The proof is obtained via straightforward application of Theorem~4 with $\hat\alpha_g$ defined by \eqref{eq-3.B-4}. 
    
    Substituting $\beta_0=\frac{1}{2\alpha_0}$ and performing simplifications yields:   
       $\hat\alpha_G=\frac{N_{IRS}}{\left(\frac{\Gamma \left(1-2 \alpha_0\right) \Gamma \left(\alpha_0\right) \Gamma \left(3 \alpha_0\right)}{\Gamma \left(1-\alpha_0\right)^2 \Gamma \left(2 \alpha_0\right)^2}\right)^2-1}$.
    Note that the term in the denominator in brackets for $\alpha_0\to 0$ equals $\frac{4}{3}$ and $2$ for $\alpha_0\to 0.316$ (see \cite[eq. 49a]{Gvo24}). Thus, $\hat\alpha_G=\frac{9 N_{IRS}}{7}$ in the first limiting case, and $\hat\alpha_G=\frac{N_{IRS}}{3}$ in the second. Using the fact that (see Theorem~\ref{thm:4}) $N_f^*$ is such a number of IRS elements for which $\hat\alpha_G=2$, and $N_n^*$ is defined when $\hat\alpha_G=\frac{1}{2}(3+\sqrt{33})$ finalizes the proof.
\end{IEEEproof}
Note that $N^*_n$ corresponds to the classical Rayleigh fading case, accounting for solution quantization (i.e., the number of elements must be integer whereas the \eqref{eq-2.B-1} solved with equality sign yields arbitrary values).

\section{Numerical example and discussion}\label{S-4}
\subsection{RIS-assisted HRR mitigation}\label{Ss-4.1}
\begin{figure}[!t]
\centerline{\includegraphics[width=\columnwidth]{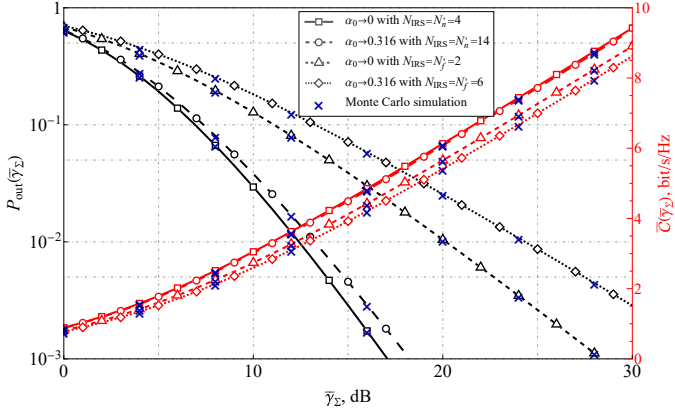}}
\caption{Exemplified IRS-assisted system performance with the results obtained from Theorem~5.}
\label{fig1}
\end{figure}

To assess the achievable gain of the quality of service (given in terms of outage probability and ergodic capacity) with the increased number of $N_{IRS}$, numerical simulations with the sample size $N_s=10^6$ were performed. Fig.~\ref{fig1} jointly illustrates the impact of the total received SNR for different values of $\alpha_0$ and $N_{\mathrm{IRS}}$ on both metrics (see the red lines for the ergodic capacity and black lines for the outage probability). The curves are obtained for the specific values $N^*_f$ and $N^*_n$ derived in Theorem~5, while the Monte-Carlo markers verify the accuracy of the proposed Gamma approximation. 
The close match between the analytical and simulated results confirms the validity of the approximation, even for small $N_{\mathrm{IRS}}$. 

\subsection{Discussion}\label{Ss-4.2}

The relations obtained in Theorem~4 between specific HRRs and channel parameters (expressed via $\hat\alpha_g$) constitute the core result, enabling the synthesis of adaptive strategies for fading mitigation through IRS element grouping. Their exemplification under particular fading conditions (Theorem~5) demonstrates that the proposed fading-centric methodology of IRS partitioning can yield relatively small configurations. This outcome can be indirectly compared with existing results. For example, the antenna-theory approach in \cite{Han21} (using the half-power beamwidth as the objective) required at least $64$ elements for rectangular IRS configurations; separate energy-efficiency and rate optimizations in \cite{Zap21} led to $100$–$200$-element arrays; whereas in \cite{Li22}, it was shown that the joint achievement of target energy efficiency and rate is possible with no more than $10$ elements (and even one in favorable conditions). Similarly, \cite{Cha25} satisfied the coverage-probability requirement with no more than $20$ elements. Hence, the exemplifying values $N_f^{*}$ and $N_n^{*}$ presented herein are fully consistent with the ranges reported in the literature.

\subsection{Limitations of the derived results and further extensions}\label{Ss-4.3}

It should be noted that although the obtained results provide a quantitative insight into HRR compensation with IRS and offer numerical estimates useful for system synthesis, they are subject to certain limitations. The mildest arise from the adopted channel model and the Gamma approximation, both well-validated in wireless studies (see~\cite{Gvo24}), and remains tight even for small $N_{\mathrm{IRS}}$. Stronger constraints stem from assuming in Theorem~2 that the $(\text{S})\!\to\!(\text{IRS}_j)$ and $(\text{IRS}_j)\!\to\!(\text{D})$ channels are independent and statistically identical for all elements~$j$ (see Theorem~3). These simplifications ensure analytical tractability but can be relaxed at the cost of heavier numerical processing. An open and important problem is the derivation of $N_f^{*}$ and $N_n^{*}$ for correlated or degenerate (pinhole) channels~\cite{Alm06}, which represents a natural direction for future research.
Moreover, it should be emphasized that the analysis was carried out for a Single-Input–Single-Output (SISO) communication system. This restriction was intentional, as it represents the worst-case scenario, given that any properly implemented diversity scheme improves link quality and reliability. Therefore, the results derived herein can be regarded as a lower-bound benchmark. Although this limitation can be relaxed by assuming diversity reception, it can be addressed either analytically -- by extending the expression for $f_{h_\Sigma}(h_\Sigma)$ using the Mellin-Barnes integral approach adopted in the proof of Theorem~2 (at the cost of cumbersome expressions) -- or by employing the closed-form or approximate models for diversity reception in the presence of generalized gamma fading derived in \cite{Cha20}.

\section{Conclusion}\label{S-5}

The presented research addresses the critical question of finding the minimum sub-array of IRS elements (i.e., how many IRS elements are needed) to mitigate severe fading conditions, quantified as hyper-Rayleigh regimes, in IRS-assisted wireless communications. By leveraging the inverse power Lomax fading model, the research derives closed-form statistics for the channel coefficient envelope and tight approximations for the total IRS-assisted link. Major findings reveal that just 2–4 IRS elements are sufficient to lift the system out of the worst-case (full-HRR) conditions, while 6–14 elements can achieve no-HRR performance, even when individual subchannels remain in full-HRR. These results provide practical guidelines for optimizing IRS sub-array sizes in 6G networks, balancing reflection efficiency and real-time adaptability.

\bibliographystyle{IEEEtran}
\bibliography{IEEEabrv,Gvozdarev_WCL2025-2703}

\begin{thebibliography}{10}
\providecommand{\url}[1]{#1}
\csname url@samestyle\endcsname
\providecommand{\newblock}{\relax}
\providecommand{\bibinfo}[2]{#2}
\providecommand{\BIBentrySTDinterwordspacing}{\spaceskip=0pt\relax}
\providecommand{\BIBentryALTinterwordstretchfactor}{4}
\providecommand{\BIBentryALTinterwordspacing}{\spaceskip=\fontdimen2\font plus
\BIBentryALTinterwordstretchfactor\fontdimen3\font minus
  \fontdimen4\font\relax}
\providecommand{\BIBforeignlanguage}[2]{{%
\expandafter\ifx\csname l@#1\endcsname\relax
\typeout{** WARNING: IEEEtran.bst: No hyphenation pattern has been}%
\typeout{** loaded for the language `#1'. Using the pattern for}%
\typeout{** the default language instead.}%
\else
\language=\csname l@#1\endcsname
\fi
#2}}
\providecommand{\BIBdecl}{\relax}
\BIBdecl

\bibitem{ITU22}
\BIBentryALTinterwordspacing
\emph{Future technology trends of terrestrial International Mobile
  Telecommunications systems towards 2030 and beyond}, ITU Recommendation
  M.2516-0, 2022-11. [Online]. Available:
  \url{https://www.itu.int/dms_pub/itu-r/opb/rep/R-REP-M.2516-2022-PDF-E.pdf}
\BIBentrySTDinterwordspacing

\bibitem{Dir19}
E.~Basar, M.~Di~Renzo, J.~De~Rosny, M.~Debbah, M.-S. Alouini, and R.~Zhang,
  ``{Wireless Communications Through Reconfigurable Intelligent Surfaces},''
  \emph{IEEE Access}, vol.~7, p. 116753–116773, 2019.

\bibitem{Zha25}
\BIBentryALTinterwordspacing
S.~Zhang, T.~Ji, M.~Hua, Y.~Huang, and L.~Yang, ``{Element-Grouping Strategy
  for Intelligent Reflecting Surface: Performance Analysis and Algorithm
  Optimization},'' 2025. [Online]. Available:
  \url{https://arxiv.org/abs/2504.15520}
\BIBentrySTDinterwordspacing

\bibitem{Zap21}
A.~Zappone, M.~Di~Renzo, X.~Xi, and M.~Debbah, ``On the optimal number of
  reflecting elements for reconfigurable intelligent surfaces,'' \emph{IEEE
  Wireless Communications Letters}, vol.~10, no.~3, pp. 464--468, 2021.

\bibitem{Li22}
D.~Li, ``{How Many Reflecting Elements Are Needed for Energy- and
  Spectral-Efficient Intelligent Reflecting Surface-Assisted Communication},''
  \emph{IEEE Transactions on Communications}, vol.~70, no.~2, pp. 1320--1331,
  2022.

\bibitem{Cha25}
Z.~Chai, J.~Xu, J.~P. Coon, and M.-S. Alouini, ``Ris-assisted millimeter wave
  communications for indoor scenarios: Modeling and coverage analysis,''
  \emph{IEEE Transactions on Vehicular Technology}, pp. 1--16, 2025.

\bibitem{Han21}
H.~Han, Y.~Liu, and L.~Zhang, ``{On Half-Power Beamwidth of Intelligent
  Reflecting Surface},'' \emph{IEEE Communications Letters}, vol.~25, no.~4,
  pp. 1333--1337, 2021.

\bibitem{Gar19}
C.~Garcia-Corrales, U.~Fernandez-Plazaola, F.~J. Canete, J.~F. Paris, and F.~J.
  Lopez-Martinez, ``{Unveiling the Hyper-Rayleigh Regime of the Fluctuating
  Two-Ray Fading Model},'' \emph{IEEE Access}, vol.~7, pp. 75\,367--75\,377,
  2019.

\bibitem{Gvo24}
A.~S. Gvozdarev, ``{Closed-Form Performance Analysis of the Inverse Power Lomax
  Fading Channel Model},'' \emph{Mathematics}, vol.~12, no.~19, 2024.

\bibitem{Gvo23}
A.~S. Gvozdarev, T.~K. Artemova, A.~M. Alishchuk, and M.~A. Kazakova,
  ``{Closed-form hyper-Rayleigh mode analysis of the fluctuating
  double-Rayleigh with line-of-sight fading channel},'' \emph{Inventions},
  vol.~8, no.~4, p.~87, 2023.

\bibitem{DLMF}
F.~W.~J. Olver, \emph{NIST handbook of mathematical functions}.\hskip 1em plus
  0.5em minus 0.4em\relax Cambridge University Press, 2010.

\bibitem{Kil04}
A.~Kilbas, \emph{{H-Transforms: Theory and Applications}}, ser. Analytical
  Methods and Special Functions.\hskip 1em plus 0.5em minus 0.4em\relax CRC
  Press, 2004.

\bibitem{Mat09}
A.~Mathai, R.~Saxena, and H.~Haubold, \emph{{The H-Function: Theory and
  Applications}}.\hskip 1em plus 0.5em minus 0.4em\relax Springer-Verlag New
  York, 2009.

\bibitem{Coh07}
A.~M. Cohen, \emph{Numerical Methods for Laplace Transform Inversion}, ser.
  Numerical Methods and Algorithms.\hskip 1em plus 0.5em minus 0.4em\relax
  Springer {US}, 2007, vol.~5.

\bibitem{Pri04}
\BIBentryALTinterwordspacing
S.~Primak, V.~Kontorovich, and V.~Lyandres, \emph{Stochastic Methods and Their
  Applications to Communications: Stochastic Differential Equations
  Approach}.\hskip 1em plus 0.5em minus 0.4em\relax Wiley, Jul. 2004. [Online].
  Available: \url{http://dx.doi.org/10.1002/0470021187}
\BIBentrySTDinterwordspacing

\bibitem{Yan20}
L.~Yang, F.~Meng, Q.~Wu, D.~B. da~Costa, and M.-S. Alouini, ``Accurate
  closed-form approximations to channel distributions of ris-aided wireless
  systems,'' \emph{IEEE Wireless Communications Letters}, vol.~9, no.~11, pp.
  1985--1989, 2020.

\bibitem{Joh94}
N.~L. Johnson, S.~Kotz, and N.~Balakrishnan,
  \emph{\BIBforeignlanguage{en}{Continuous Univariate Distributions, Volume
  1}}, 2nd~ed., ser. Wiley Series in Probability and Statistics.\hskip 1em plus
  0.5em minus 0.4em\relax Nashville, TN: John Wiley \& Sons, Oct. 1994.

\bibitem{Alm06}
P.~Almers, F.~Tufvesson, and A.~F. Molisch, ``{Keyhole Effect in MIMO Wireless
  Channels: Measurements and Theory},'' \emph{IEEE Transactions on Wireless
  Communications}, vol.~5, no.~12, pp. 3596--3604, 2006.

\bibitem{Cha20}
T.~Chaayra, H.~Ben-azza, and F.~E. Bouanani, ``{New Accurate Approximation for
  the Sum of Generalized Gamma Distributions and its Applications},''
  \emph{Applied Mathematics $\&$ Information Sciences}, vol.~14, no.~6, p.
  1121–1136, 11 2020.

\end{thebibliography}

\end{document}